\documentclass[aps,prd,10pt,onecolumn,nofootinbib,groupedaddress,amsfonts,floatfix,notitlepage]{revtex4-1}

\usepackage[colorlinks=true,urlcolor=blue,linkcolor=,citecolor=blue]{hyperref}
\usepackage{amsmath,amssymb,amstext,amsbsy,amsfonts,amsthm,graphicx,microtype,dsfont,verbatim}
\usepackage[capitalize]{cleveref}
\usepackage{ulem}
\usepackage[dvipsnames]{xcolor}

\creflabelformat{equation}{(#2\textup{#1}#3)}

\newcommand{\prg}{\mathrm{pr}}
\newcommand{\mpl}{M_\mathrm{Pl}}

\newcommand{\diff}[1]{\mathrm{d}{#1}\,}

\def\be{\begin{equation}}
\def\ee{\end{equation}}
\def\ba{\begin{eqnarray}}
\def\ea{\end{eqnarray}}

\renewcommand{\d}{\mathrm{d}}

\begin{document}

\title{Simulating a numerical UV Completion of Quartic Galileons}

\author{Mary Gerhardinger${}^1$}
\author{John T. Giblin, Jr${}^{2,3,4}$}
\author{Andrew J. Tolley${}^{5,3}$}
\author{Mark Trodden${}^1$}
\affiliation{${}^1$Center for Particle Cosmology, Department of Physics and Astronomy, University of Pennsylvania, Philadelphia, Pennsylvania 19104, USA}
\affiliation{${}^2$Department of Physics, Kenyon College, Gambier, Ohio 43022, USA}
\affiliation{${}^3$CERCA/ISO, Department of Physics, Case Western Reserve University, Cleveland, Ohio 44106, USA}
\affiliation{${}^4$Center for Cosmology and AstroParticle Physics (CCAPP) and Department of Physics, The Ohio State University, Columbus, OH 43210, USA}
\affiliation{${}^5$Theoretical Physics, Blackett Laboratory, Imperial College, London, SW7 2AZ, UK}

\email{maryge@sas.upenn.edu \\ giblinj@kenyon.edu \\ a.tolley@imperial.ac.uk \\ trodden@physics.upenn.edu}

\begin{abstract}
The Galileon theory is a prototypical effective field theory that incorporates the Vainshtein screening mechanism---a feature that arises in some extensions of General Relativity, such as massive gravity. The Vainshtein effect requires that the theory contain higher order derivative interactions, which results in Galileons, and theories like them, failing to be technically well-posed.
While this is not a fundamental issue when the theory is correctly treated as an effective field theory, it nevertheless poses significant practical problems when numerically simulating this model. These problems can be tamed using a number of different approaches: introducing an active low-pass filter and/or constructing a UV completion at the level of the equations of motion, which controls the high momentum modes. These methods have been tested on cubic Galileon interactions, and have been shown to reproduce the correct low-energy behavior.  Here we show how the numerical UV-completion method can be applied to quartic Galileon interactions, and present the first simulations of the quartic Galileon model using this technique. We demonstrate that our approach can probe physics in the regime of the effective field theory in which the quartic term dominates, while successfully reproducing the known results for cubic interactions. 

\end{abstract}

\maketitle


\section{Introduction}
While General Relativity (GR) has made myriad successful predictions at many different scales, a number of deep puzzles posed by contemporary cosmology have led to great interest in the question of whether the theory might admit robust modifications. The effective field theory formalism has provided a particularly useful way of exploring and categorizing such modifications, while ensuring that GR is recovered in the local regimes where it has been so stunningly successful.  
Modifications of gravity and general models of dark energy and cosmic acceleration can be characterized by the mechanisms through which they reproduce local scale gravity---i.e. which screening mechanism they employ. Of particular interest in this paper are those theories that use the Vainshtein screening mechanism \cite{Vainshtein:1972sx} to suppress deviations from GR by screening out the fifth-force on local scales \cite{Deffayet:2001uk,Dvali:2002vf,Lue:2003ky,Babichev:2009us,Babichev:2009jt,Babichev:2010jd,deRham:2010ik,deRham:2010kj,Babichev:2013usa,Joyce:2014kja}. 
There exists a range of complicated theories which exhibit the Vainshtein mechanism, but the Galileon~\cite{Nicolis:2008in}, a model with a single real scalar field, is the simplest example that encapsulates all the essential features, and is known to arise naturally in the decoupling limit of massive gravity theories \cite{Luty:2003vm,Babichev:2009us,Babichev:2009jt,Babichev:2010jd,deRham:2010ik,deRham:2010kj,Ondo:2013wka,deRham:2023byw}. \\

Galileons include higher order derivative interaction terms which give rise to the Vainshtein screening mechanism \cite{Babichev:2009us,Babichev:2009jt,Babichev:2010jd} yet they have second order equations of motion.
Away from extremely special configurations, such as spherically symmetric, static sources, Galileons are difficult to describe analytically, precisely because of these nonlinear interactions. Hence numerical investigations are critical for understanding this model. Although such studies are made easier by the second order nature of the equations of motion, the effective theory of the Galileon nevertheless remains not well-posed; it has regimes in which the equations of motion are non-hyperbolic. These regimes arise when the nonlinear interactions are very large and the effective field theory is no longer necessarily under control. While this does not impact the theoretical underpinnings of the model---it is entirely well-behaved when properly interpreted as an effective field theory---in order to successfully describe the dynamics of the Galileon system, numerical simulations need to avoid or mitigate these unstable regions. \\

The cubic Galileon model was considered in \cite{Dar:2018dra} and it was shown that by turning the nonlinear interactions on slowly it was possible to mitigate any potential numerical instabilities, and obtain a power spectrum which matched the general scaling expectations of analytic estimates \cite{deRham:2012fw,Chu:2012kz,deRham:2012fg}. This was significantly improved upon in  \cite{Gerhardinger:2022bcw} where we considered the numerical evolution of the cubic Galileon using three distinct methods. The first amounted to low-pass filter combined with the slow turn off method of \cite{Dar:2018dra}. The other two methods involved replacing the Galileon theory with a system of auxiliary higher spin fields that control the high frequency modes---a well-posed (numerical) UV completion model. 
Motivated by the origins of the Galileon as a massive spin 2 field, these additional spin fields propagate via either hyperbolic or parabolic equations of motion. 
The parabolic model is similar to systems proposed in \cite{Cayuso:2017iqc,Allwright:2018rut} that have been successfully applied to simulating evolution within scalar-Gauss-Bonnet gravity \cite{Franchini:2022ukz,Thaalba:2023fmq},  \textit{k}-essence \cite{Bezares:2021yek,Lara:2021piy,Barausse:2022rvg,Boskovic:2023dqk}, Horndeski gravity \cite{Ripley:2022cdh}, and other extensions to General Relativity \cite{Cayuso:2020lca,Cayuso:2023aht,Cayuso:2023xbc}.
This technique is predicated on the idea that the theory of interest is a truncation of some larger theory that is well-behaved, and the problems arise from the truncation \cite{Franchini:2022ukz}.
These approaches, often referred to as \textit{fixing-the-equations}, are based on the M\"uller-Israel-Stewart formulation \cite{muller1967paradoxon,ISRAEL1976213,ISRAEL1979341,ISRAEL1976310}, in which the additional degrees of freedom obey their own wave equations. \\

In our work on the cubic Galileon \cite{Gerhardinger:2022bcw}, we demonstrated that the UV completion model asymptotically approaches the original cubic Galileon theory in the low energy limit, and argued that our numerical treatment correctly reproduces the dynamics of the Galileon in the IR regime. It is worth emphasizing that the system provides a purely numerical completion, as there is no known Lorentz invariant local and unitary UV completion of the Galileon, and indeed there are suggestions that such a theory does not exist \cite{Tolley:2020gtv}. We also compared the three different ways of resolving the physics in the cubic Galileon model with numerical integration techniques.  All three models reproduced the same long wavelength physical processes up to expected numerical errors. \\


In this paper we take the next important step to understanding the Vainshtein screening by including both cubic and quartic Galileon interactions. This is by no means straightforward since it is known that despite being moderately successful for the cubic Galileon \cite{deRham:2012fw,Dar:2018dra}, analytic attempts to describe the power radiated from a rotating system completely fail in the case of the quartic Galileon \cite{deRham:2012fg}. This occurs because the large multipoles are not sufficiently suppressed. For this reason it has remained unclear if the Vainshtein mechanism is even active for a time-dependent system when the quartic Galileon is active.
In the present paper, we extend our successful discussion of the cubic Galileon using a well-posed UV completion of the equations of motion to account for quartic interactions. In a companion paper we similarly extend the low-pass filter method to the quartic case \cite{Giblin:2024}.  \\


\section{Quartic Galileon}
The Galileon is a real scalar field, $\pi$, which satisfies the symmetry $\pi \rightarrow \pi +b_{\mu}x^{\mu} +c$, for constant parameters $b_{\mu}$ and $c$. The relevant action for the Galileon including both cubic and quartic interactions is  \cite{Nicolis:2008in,Joyce:2014kja,deRham:2012fg}) is\footnote{Throughout, we work in the mostly positive metric convention. }
	\begin{equation}
		\label{eqn:galaction}
		S = \int \diff{^4 x}\left(-\frac{3}{4} (\partial\pi)^2 - \frac{1}{4\Lambda_3^3} (\partial\pi)^2 \Box \pi\\ 
		- \frac{1}{24\Lambda_4^6}(\partial\pi)^2\left((\Box\pi)^2-(\partial\partial\pi)^2\right) + \frac{1}{2\mpl}\pi T \right) \ ,
	\end{equation}
where $\Lambda_3$ and $\Lambda_4$ control the strengths of the nonlinear couplings in the cubic and quartic terms respectively and $T$ is the trace of the stress energy tensor for the matter content of the theory. The slightly unusual choice of normalization is a reflection of how $\pi$ emerges as the helicity-zero mode in massive gravity theories where it naturally couples to the trace of the stress energy momentum tensor. This gives rise to a classical equation of motion
	\begin{equation}
	\label{eqn:eom}
		 \Box\pi + \frac{1}{3\Lambda_3^3}\left((\Box\pi)^2 - (\partial_\mu \partial_\nu \pi)^2 \right) + \frac{1}{9\Lambda_4^6} \left((\Box\pi)^3 - 3\Box\pi (\partial_\mu \partial_\nu \pi)^2 + 2 (\partial_\mu \partial_\nu \pi)^3 \right) = \frac{T}{3\mpl} \ ,
	\end{equation}
with $(\partial_\mu \partial_\nu \pi)^3 \equiv (\partial_\mu \partial^\alpha \pi)(\partial_\alpha  \partial^\nu \pi)(\partial_\nu \partial^\mu \pi)$.  
The most important type of nonlinearity is controlled by the relative magnitudes of $1/ \Lambda_3$ and $1/\Lambda_4$. 
For a spherically symmetric, time-independent source with $T = \rho(r)$ the Galileon system is well studied~\cite{Nicolis:2004qq,deRham:2012fg,Berezhiani:2013dca}, and the equation of motion reduces to
	\begin{equation}
		\frac{1}{r^2} \frac{d}{dr} \left[ r^3 \left(  \frac{E(r)}{r}+ \frac{2}{3 \Lambda_3^3} \left(\frac{E(r)}{r} \right)^2 + \frac{2}{9 \Lambda_4^6 }\left(\frac{E(r)}{r}\right)^3  \right)\right] = \frac{\rho(r)}{3 \mpl} ,	\end{equation}
where $E(r) \equiv \partial \pi / \partial r$. We can integrate this to obtain a solution of the form
	\begin{equation}
		\frac{E(r)}{r} + \frac{2}{3 \Lambda_3^3} \left(\frac{E(r)}{r} \right)^2 + \frac{2}{9 \Lambda_4^6 }\left(\frac{E(r)}{r}\right)^3 = \frac{M_{s}(r)}{12 \pi  \mpl} \frac{1}{r^3} \ ,
	\end{equation}
	where 
	\be
	M_s(r) = 4 \pi \int_0^{r} \d r \, r^2 \rho(r) ,
	\ee
	is the mass contained within a finite radius when the density is approximated as a function of $r$ only. Well outside the source we may take $M_s(r)=M_s$ constant, without considering higher moments of $\rho$. This solution reveals two important distances associated with the two scales: $\Lambda_3$ and $\Lambda_4$.
We define the {\it Vainshtein radii} of the source as the radial locations at which the different nonlinear interaction terms become important, given by
	\begin{align}
		& r_{*,3} \equiv \left( \frac{M_{s}}{16 \mpl} \right)^{1/3} \frac{1}{\Lambda_3} \, ,  \\ \label{eq:screen3} 
		& r_{*,4} \equiv  \left( \frac{M_{s}}{16 \mpl} \right)^{1/3} \left( \frac{\Lambda_3}{\Lambda_4}\right)^3 \frac{1}{\Lambda_4} \ ,
	\end{align}
where $r_{*,3}$ is the radius at which the cubic term is approximately the same size as the kinetic term, and $r_{*,4}$ is that at which the quartic term is approximately the same size at the cubic term \cite{deRham:2012fg}.
The Vainshtein radii define the distances from the source at which the nonlinear interaction terms become important. It is worth noting that for astrophysical sources, and cosmologically-interesting cutoffs, the Vainshtein radii are extremely large distances. \\

For more complicated sources, the Galileon system has been studied both analytically and numerically (see e.g. \cite{deRham:2012fg,deRham:2012fw,Chu:2012kz,deRham:2012az,Dar:2018dra,Kaloper:2011qc,Iorio:2012pv,Barreira:2013eea,Ogawa:2018srw}.
Numerical challenges arise when simulating realistic sources in this model --- see for instance \cite{Allwright:2018rut,Cayuso:2023xbc}. 
Well-posed field theories describe physics at IR energy scales that are decoupled from the UV modes of the theory. However truncated EFT expansions are generically not well-posed. Although not a fundamental issue (for example there is no ambiguity in inferring low energy S-matrix elements) this can create a significant problem for the numerical evolution of the classical system since the UV and IR modes might be coupled to one another  \cite{Solomon:2017nlh}. Furthermore when the interactions are higher derivative, as is generic, the evolution may transfer energy from the IR into the UV modes leading to unstable solutions.
The question surrounding the well-posedness problem, then, is if it is possible to manage the UV modes without affecting the physics occurring in the IR region. \\

In our previous paper \cite{Gerhardinger:2022bcw}, we introduced and verified two classes of techniques which can accomplish this: using a low-pass filter to dampen high frequency modes, and introducing additional fields to implement a UV completion at the level of the equations of motion. Simulations of the cubic Galileon involving each of these techniques produced results that were both stable in evolution and reproduced solutions consistent with analytical expectations. 
In what follows we extend the second approach to the Galileon system with both cubic and quartic interactions.


\subsection{The numerical UV Completion}

Following  \cite{Gerhardinger:2022bcw} our goal is to introduce a numerical UV completion which is designed to control the high energy behavior and allow for more stable numerical evolution than the original Galileon system. 
Note that this is not a true UV completion, i.e we do not require an action, rather this is a device to regulate the numerical evolution. 
With this in mind, we are allowed to introduce friction terms that mildly break Lorentz invariance to help prevent the instabilities of high $k$ modes \cite{Keltner:2015xda}. 
Using the same technique as in the cubic Galileon case \cite{Gerhardinger:2022bcw}, 
inspired by how the Galileon can arise as the helicity zero mode of a spin 2 field within massive gravity theories, we introduce an auxiliary massive spin 1 $A_\mu$ and a massive spin 2 $H_{\mu \nu}$ field that satisfy damped harmonic oscillator equations \cite{Arkani-Hamed:2002bjr,deRham:2010ik,deRham:2010kj,deRham:2014zqa}. 
These new fields replace the troublesome higher order derivative terms in the Galileon equation of motion with functions of the $H_{\mu \nu}$ fields; thereby allowing us to control the derivative interaction terms that are the origin of  the instability problems. 
We first define auxiliary fields via
	\begin{equation}
	\label{eqn:AandHDefs}
		A_{\mu} \equiv \partial_{\mu} \pi ,
\end{equation}
and
\begin{equation}
H_{\mu\nu} \equiv \frac{1}{2} ( \partial_{\mu} A_{\nu}+ \partial_{\mu} A_{\nu}) .
\end{equation}
	We then promote the auxiliary fields to have their own dynamics and, at the same time, trade the higher derivative Galileon interactions for the lower derivative fields so that the UV theory is described by the equations of motion
    \begin{eqnarray}
    \label{eqn:pieom2}
	&& \ddot{\pi} = \triangledown^{2}\pi + \frac{1}{3\Lambda_3^3}\left( (H_{\nu}^{\nu})^2 - H_{\mu\nu} H^{\mu\nu} \right) +  \frac{1}{9\Lambda_4^6}\ \left( (H^{\nu}_{\nu})^3 - 3 H^\alpha_\alpha H_{\mu \nu} H^{\mu \nu} + 2 (H_{\mu \nu})^3 \right) - \frac{T}{3 \mpl} \, , \\
    \label{eqn:Aeom}
	&&\ddot{A}_{\mu} = \triangledown^2 A_{\mu} - \frac{1}{\tau} \partial_{0} A_{\mu} - M^2 ( A_{\mu} - \partial_{\mu} \pi ) \, , \\
     \label{eqn:Heom}
	&&\ddot{H}_{\mu\nu} = \triangledown^2 H_{\mu \nu} - \frac{1}{\tau} \partial_0 H_{\mu\nu} - M^2 ( H_{\mu\nu} - \frac{1}{2} (\partial_{\mu} A_{\nu} + \partial_{\nu} A_{\mu})) \ . 
	\end{eqnarray} 
These equations have been constructed such that the friction term ensures that the new propagating degrees of freedom in $A_\mu$ and $H_{\mu \nu}$ decay on a time order of $\tau$.
As in the cubic case, we have now turned a single propagating field into a total of fifteen fields, all of which obey a form of the wave equation. Importantly, the fourteen additional degrees of freedom should decay away on a time scale of order $\tau$.  
The only difference between this quartic Galileon system and the cubic Galileon system considered in \cite{Gerhardinger:2022bcw} is an additional term in the equation of motion for the $\pi$ field. \\

The terms on the right-hand side of \eqref{eqn:Aeom} and \eqref{eqn:Heom} dictate that solutions will asymptote at low energies ($k, \omega \ll M$) to the definitions of $A_\mu$ and $H_{\mu \nu}$ respectively, given in \eqref{eqn:AandHDefs}. 
Additionally, the definitions of the auxiliary fields are constraints on the boundary conditions of our numerical system.
In the low energy limit, it is clear to see that \eqref{eqn:pieom2} reduces exactly to \eqref{eqn:eom}, the equation of motion for the Galileon that we started with.
The addition of these fields does not guarantee stability, however it does eliminate any issues that arise from the derivative interaction terms.  
For another way to write out this numerical UV completion---one which makes the connection to the IR regime more apparent---see \cite{Gerhardinger:2022bcw}.


\section{Numerical Simulations}
\label{sec:NumSims}
The Galileon effective field theory poses several problems for numerical simulations due to its nonlinear nature, coupled derivative interaction terms, and the many scales of interest present in the system\footnote{Other work has focused on singularities in the effective metric of perturbations as a cause for numerical issues \cite{Brito:2014ifa}.}. 
Having shown in a previous work that the full cubic system can be solved numerically in a way that produces results which agree with analytic expectations, we use here the most stable of our methods to produce the simulations of the quartic Galileon model. These numerical simulations are conducted using GABE, a verified numerical program for simulating scalar fields \cite{Child:2013ria}.\footnote{http://cosmo.kenyon.edu/gabe.html}  \\

To be concrete, we focus on the case when the source is comprised of two rotating Gaussian-shaped mass-energy distributions. This is a challenging example that is closely related to physically relevant sources and also allows for easy comparisons with our previous work with the cubic system as well as with other numerical simulations \cite{Dar:2018dra} and analytic expectations \cite{deRham:2012fw,Chu:2012kz}.
We parameterize the system through two dimensionless quantities: $\alpha \equiv \Omega \bar{r}$, which describes the rotational speed of the system, and  $\beta \equiv \bar{r}/r_{*,3}$, which relates the diameter of the sources to the cubic Vainshtein radius as defined by \cref{eq:screen3}. 
To fully constrain the system, we can use Kepler's Law (a reasonable approximation for small velocities)
	\begin{equation}
		\Omega^2 = \frac{M_s}{8 \pi \mpl^2 \bar{r}^3} ,
	\end{equation}
where $M_s$ is the total mass of the system, to describe the cubic nonlinear coupling term via
	\begin{equation}
		\kappa_{\text{cubic}} = \frac{32}{3 \sqrt{2 \pi}} \beta^{-3} \alpha^{-1}.
	\end{equation}
We choose $\beta = 0.05$ and $\alpha = 0.2$ as our fiducial model, which translates into a coupling strength $\kappa_{\text{cubic}} \approx 1.7 \times 10^5.$

\subsection{Dimensionless Units}
\label{sec:DimUnits}
For the simulation to be independent of the physical parameters of this model, we rescale both our scalar field and the spatial coordinates to yield dimensionless variables (denoted with the subscript $\prg$), defined by

\begin{equation}		\pi_{\prg} = \pi \sqrt{\frac{\bar{r}}{M_s}} , \quad   x^{\mu} = \frac{\bar{r}}{2} \, ,
   \end{equation}
where $\bar{r}$ is the distance between the centers of the rotating Gaussians, and the rescaling of the spatial derivative clearly follows from this latter definition.
These allow us to rewrite the equation of motion given in \eqref{eqn:eom} as
	\begin{equation*}
		\square_{\prg} \pi_{\prg} + \kappa_{\text{cubic}} \left((\square_{\prg} \pi_{\prg})^2 - (\partial_\mu^{\prg} \partial_\nu^{\prg} \pi_{\prg})^2 \right)  + \xi^6 \kappa_{\text{cubic}}^2 \Big( (\square_{\prg} \pi_{\prg} )^3 - 3 \square_{\prg} \pi_{\prg} (\partial_\mu^{\prg} \partial_\nu^{\prg} \pi_{\prg})^2 
		 + 2 (\partial_\mu^{\prg} \partial_\nu^{\prg} \pi_{\prg})^3 \Big) = J_{\prg},
	\end{equation*}
where we regularize the sources using $J_{\prg}$
\begin{equation}
\label{eqn:source}
        J_\prg = A \left(e^{-\left({\vec{r}_+}^{\,\prg}(t)/\sigma_\prg\right)^2}+e^{-\left({\vec{r}_-}^{\,\prg}(t)/\sigma_\prg\right)^2}\right);
    \end{equation}
where $ {\vec{r}_\pm}^{\,\prg} (t) = \left(x_\prg\pm \cos\left(\Omega_\prg t_\prg\right), y_\prg \pm \sin\left(\Omega_\prg t_\prg\right), z_\prg\right)$ and the constant 
\begin{equation}
A \equiv \frac{2\sqrt{2}}{3\pi} \frac{\Omega \bar{r}}{\sigma_\prg^3} ,
\end{equation}
are chosen to ensure that the total mass of the system is given by $M_{\rm s} = \int d^3x\, \rho = \int d^3 x \,T$, as expected. 
Additionally, we define the dimensionless quantities 
	 \begin{equation}
  	 \label{eqn:kappaquartic}
        		\kappa_{\text{cubic}} = \frac{1}{3 \Lambda_3^3} \sqrt{ \frac{16 M_s}{\bar{r}^5} }  \quad \text{ and } \quad \xi \equiv \frac{\Lambda_3}{\Lambda_4},
    	\end{equation}
 which control the size of the nonlinear terms and the strength of the quartic term relative to the cubic term, respectively.
In the limit $\xi \rightarrow 0$, this system will reduce to the cubic Galileon model. Additionally, the auxiliary spin fields are rescaled according to 
	\begin{equation}
   	     A_\mu = \sqrt{\frac{4 M_s}{\bar{r}^3}} {A_\mu}^{\prg} \, , \quad
	     H_{\mu \nu} = \sqrt{\frac{16M_s}{\bar{r}^5}} {H_{\mu\nu}}^{\prg} .
 	   \end{equation}
We investigate the quartic Galileon system, with these sources, in a variety of regimes; we also define a fiducial case for which the physical and numerical parameters are: box size $L = 2.5 r_{*,3} = 50 \bar{r}$, number of points along the box $N = 384$, and time step size $dt = \beta^{-1}\bar{r} /  6400 = 0.003125\bar{r}.$

\subsection{Analytic Expectations}
\label{sect:analytics}
\begin{figure*}[tbp]
\centering
\includegraphics[width=0.5\textwidth]{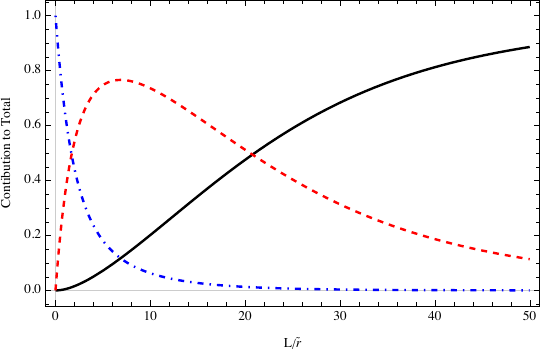}
\caption{Analytic contribution of each of the Klein-Gordon (solid black line), cubic (red dashed), and quartic (blue dot-dashed) terms to the total $\pi$ field for our fiducial model with $\alpha = 0.2$, $\beta = 0.05$ and $\xi = 0.6$ illustrating the two distinct Vainshtein screening regimes.  \label{AnalyticExpec}}
 \end{figure*}

The analytic solution for a single Gaussian source is determined by the polynomial equation
	\begin{equation}
	r^2 \mathcal{E}(r) + 2 r \kappa_{\text{cubic}} \mathcal{E}(r)^2 + 2 \kappa_{\text{cubic}}^2 \xi^6 \mathcal{E}(r)^3 =\alpha \sqrt{\frac{2}{9 \pi}},
	\end{equation}
where we define $\mathcal{E}(r) \equiv \partial \pi_\prg / \partial r_\prg$. It is evident that as  $\xi \rightarrow 0$ we recover the cubic system. For $\xi>0$ there will be 3 regions depending on which of the three terms Klein-Gordon, cubic, and quartic dominates. We can solve this analytically and determine the relative importance of each of the three terms for a given value of $\xi$.
Figure \ref{AnalyticExpec} depicts the contribution of each of these terms to the equation of motion for the $\pi$ field for a given value of $\xi = 0.6$ and a single Gaussian source. 
It is clear that for $\xi=0.6$, the binary system is well inside the region in which the quartic Galileon dominates, namely the Vainshtein radius $r_{*,4}$. For this reason we expect our simulations to be substantively different than those for the pure cubic Galileon $\xi=0$.

 \subsection{Boundary Conditions}
In each simulation, we use outgoing boundary conditions, so that the physical processes inside the box are the only sources to the fields.
To isolate the system, we construct a buffer of points inside the lattice at each boundary where the field is no longer propagating according to its equation of motion given in \eqref{eqn:pieom2} but rather evolves using outgoing boundary conditions. 
For the scalar $\pi$ field---or any interaction-less and massless field---this looks like 
	\begin{equation}
		\dot{\pi} = -\frac{\pi}{r} - \partial_r \pi.
	\end{equation}
Although this works well for the $\pi$ field, problems arise for the massive auxiliary fields. 
As in the cubic case, we minimize those issues by enforcing a parabolic damped constraint equation, given by
	\begin{eqnarray}
		\dot{A}_\mu &=& - C (A_\mu - \partial_\mu \pi)\\
		\dot{H}_{\mu \nu} &=& - C \left(H_{\mu \nu} - \frac{1}{2}\left( \partial_\mu A_\nu + \partial_\nu A_\mu \right)\right),
	\end{eqnarray}
where $C$ is some parameter that determines the extent to which the auxiliary fields are damped.
For our simulations, we set this constant by the decay scale for the auxiliary fields, $C = M_{\prg}^2 \tau$, a parameter to which the simulations are largely insensitive below a value of $C = \tau \times 10^4$. 
These are the same `damped' versions of the constraint equations used in the cubic system \cite{Gerhardinger:2022bcw}, where we determined that directly enforcing the constraint equations on the boundary introduces large discontinuities that caused backreaction into the box. 
In the low energy limit, we expect that $A_\mu \sim \partial_\mu \pi$ and $H_{\mu \nu} \sim \partial_\mu \partial_\nu \pi$, which would result in a zero time derivative---and no backreaction. \\

As energy increases beyond the low energy limit, these approximations no longer hold, and the time derivatives of the auxiliary fields are no longer trivial.
In that case, our goal is to minimize the effect that the discontinuities have on the system, and to enforce that the auxiliary fields behave like derivatives of the $\pi$ fields, so that discontinuities do not disrupt the physical processes occurring at low energies. 
The discontinuities appear because of the contribution of the cubic terms to the equation of motion at the edge of the box. 
At the boundary, we assume that the cubic terms vanish, but as Figure \ref{AnalyticExpec} depicts, there is still around a 40 \%  contribution at the boundary. 
At first glance, one viable solution to this appears to be to make the box bigger in order to allow the cubic contribution terms to vanish more completely before reaching the boundary. 
However this significantly reduces the resolution of the sources in the middle of the box. 
Hence, we enforce damped constraint equations to mitigate backreaction from the discontinuities and to maximize the spatial resolution of the source.

 \subsection{Power Calculations}
 \label{sect:power}
To establish whether the Vainshtein mechanism is active for a time-dependent rotating binary source with the quartic interactions present we can calculate the power emitted per multipole. 
We calculate this in the same way as described in \cite{Gerhardinger:2022bcw}, by setting up a sphere of points, evaluating $\pi$, $\dot{\pi}$, and $\partial_r \pi$ at those points, and then performing a tri-linear interpolation to project the field onto the spherical harmonics.
The sphere is set up using the {\sc HEALPIX}\footnote{http://healpix.sourceforge.net} standard, defined by a radius of $r = 22.5 \bar{r}$ which is larger than the Vainshtein radius, $r_v = 20 \bar{r}$, but smaller than half the size of the box, $L_{\text{half}}  = 25 \bar{r}$. 
This method provides us with a sphere of evenly spaced points (not necessarily at the lattice point locations) which are used to decompose the field onto spherical harmonics. \\

In \cite{deRham:2012fg}, the authors analytically found that the power in higher multipoles are suppressed and that the dominant multipole was the quadrupole. This was confirmed by the numerical simulations of \cite{Dar:2018dra,Gerhardinger:2022bcw}. By contrast in \cite{deRham:2012fw} it was found that including a quartic interaction term greatly alters the behavior of perturbations around a spherically symmetric, time dependent source rendering the approximate perturbative treatment inadequate for calculating the radiated power. Nevertheless some insight can be gained from considering the form of the analytic radiated power which should be valid when there is a large hierarchy between the scales involved which consequently allows for the linear treatment of perturbations from a static, spherically symmetric background solution. In this situation, the radiated power for a given mode depends on $m$ and $\ell$, derived in Appendix A of \cite{Dar:2018dra}, and is given by 
	\begin{equation}
		P_\ell = \frac{\pi \Omega_p M^2}{12 \mpl^2} \sum^{\ell}_{m=0} m (1+ (-1)^m) u_{\ell m}^2 (\bar{r} / 2) | Y_{\ell m} (\pi / 2,0) |^2 \ ,
	\end{equation}
where we have restricted ourselves to circular orbits in the $\theta = \pi/2$ plane with equal mass objects, 
For $\ell = 0$, we are constrained to $m = 0$, which causes the power to vanish, as we expect for the monopole.
For $\ell = 1$, the two allowed values for $m$ also lead to zero power: $\ell = 1, m = 0$ is zero because of the $m$ contribution at leading order, $\ell = 1, m = 1$ is zero because the $(1 + (-1)^m)$ term is also zero. 
For $\ell = 2$, the first non vanishing power term, $m = 2$ is the sole contributor because $m$ must be \textgreater 0, and it must be even, so $1 + (-1)^m \neq 0$.
For $\ell = 3$, the constraints on $m$ leave only the $m = 2$ case as a possibility for non-zero power. However, this is also zero because $Y_{\ell=3,m=2}(\pi/2,0)$ vanishes.
Extrapolating, we expect that the only non-vanishing power will arise in modes that have even $\ell$ ($\ell = 2n \text{ where } n \in \mathds{N}$) with contributions from $m$ modes that are also even and greater than zero. 


\subsection{Full Auxiliary Field Method}
In our previous work, we described two distinct numerical UV completions for simulating the cubic Galileon model, both of which reproduced the long wavelength behavior expected by analytic analysis. 
While each of these had their own advantages and shortcomings, we have chosen to employ the `Full Auxiliary Field' method for simulating the quartic system because it allows nonlinear interactions to turn on sooner and generally requires less spatial resolution. 
Our chosen UV completion is in program units 
	\begin{align}
		&\Box_\prg \pi^\prg + \kappa \left(H_\prg^{\mu \nu}H^\prg_{\mu \nu} - \left({H_\prg}^\nu_\nu\right)^2\right) = f_1(t) J_\prg \ , \label{Piequationnum}\\
		&\Box_\prg A^\prg_\mu - \frac{1}{\tau} \partial^\prg_t A^\prg_\mu - M_\prg^2 A^\prg_\mu = - M_\prg^2 \partial^\prg_\mu \pi^\prg \ , \label{Aequationnum}\\
		&\Box_\prg H^\prg_{\mu\nu} - \frac{1}{\tau} \partial^\prg_t H^\prg_{\mu\nu} - M_\prg^2 H^\prg_{\mu\nu} = - \frac{M_\prg^2}{2}\left(\partial^\prg_\mu A^\prg_\nu + \partial^\prg_\nu 			A^\prg_\mu\right) \, , \label{Hequationnum}
	\end{align}
where $M_\prg = M \bar{r}/2$ and $f_1(t)$ is a window function given by
	\begin{equation}
		f_1(t_{\prg}) = \frac{1}{2}\tanh^{-1}\left(\frac{1}{10}\left[t_{\prg}-25\right]\right)+\frac{1}{2} \, ,
	\end{equation}
that ramps up from zero to unity.  
Using this system and the relaxed numerical constraints as boundary conditions for the auxiliary fields, we have been able to achieve numerically stable simulations of the quartic Galileon model. \\

\begin{figure*}[tbp]
\centering
\includegraphics[width=0.5\textwidth]{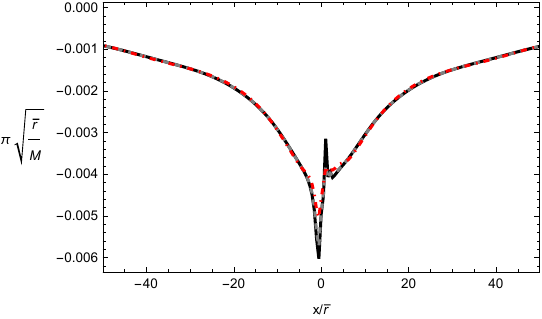}
\caption{(Left) The $\pi$ field profile along the $x$ axis after the system has reached stability using a mass value of $M_{\prg} = 3$ and an increasing nonlinear quartic term, controlled by $\xi$. The cubic case ($\xi = 0$) is denoted by an solid black line, the $\xi = 0.4$ a gray dashed line, and $\xi = 0.6$ a red dot-dashed line.
\label{fieldprofile}}
 \end{figure*}

An important check on the numerics is to first ensure that the quartic simulations reproduce the cubic system previously simulated by taking the limit $\xi \rightarrow 0$ for a given mass value. 
Figure \ref{fieldprofile} depicts the asymptotic approach of the $\pi$ field profile from the quartic system toward the field profile of the cubic system, taken along a line in the equatorial plane of the system.
We argued in \cite{Gerhardinger:2022bcw} that the Full Auxiliary field method reproduces the dynamics of the cubic system in a way that matches analytic expectations---here we validate the current method by comparing it to previous cubic work \cite{Gerhardinger:2022bcw}. \\

\begin{figure*}[tbp]
\centering
\includegraphics[width=0.5\textwidth]{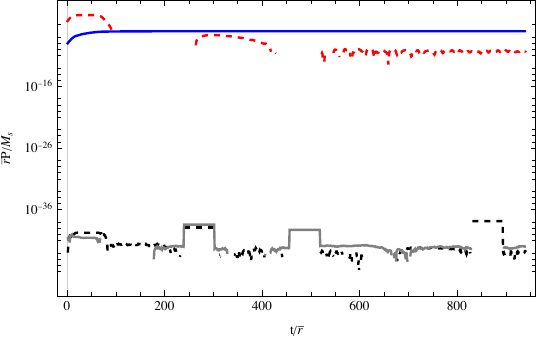}
\caption{ Instantaneous and averaged power emitted in the fiducial system with $M\bar{r} = 3.76$ and $\xi = 0.8$ for the monopole up to $\ell = 3$ modes clearly demonstrating the dominance of the quadrupole at late times. The Monopole is denoted by a red, dashed line, the dipole a black dashed line, the quadrupole a blue solid line, and the $\ell = 3$ pole a solid gray line. The dipole and $\ell = 3$ modes are both at machine zero. 
 \label{PowersOfXi}}
 \end{figure*}

In addition, we have investigated the dynamics of this quartic model by examining the multipole power in the system radiated by the $\pi$ field.
Figure \ref{PowersOfXi} depicts the period-averaged power contained in each multipole for a single value of $\xi$, the parameter controlling the relative strength of the quartic term. The multipoles diminish in power as the multipole number increases and only select even poles ($\ell = 2n, m = 2n \text{ where } n \in \mathds{N}$) have appreciable nonzero power, with any power in odd multipoles arising from numerical errors. Although these results match the analytic expectations discussed in Section \ref{sect:power} we stress that for comparable mass binary sources for which the orbit lies inside the quartic region, the analytic approach breaks down. Our result is hence nontrivial and match the results of the low-pass filter method in \cite{Giblin:2024}. \\

\begin{figure*}[tbp]
\centering
\includegraphics[width=0.5\textwidth]{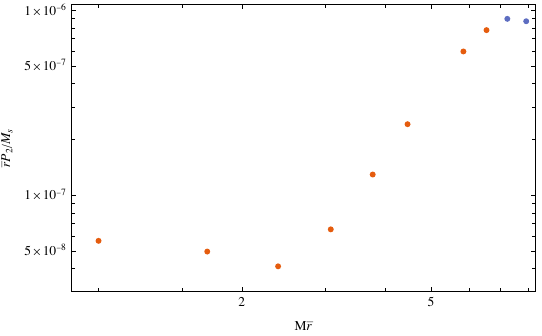}
\caption{Late-time quadrupole power emitted by the fiducial system with a constant value of $\xi = 0.6$ and changing $M$ parameter. The orange dots have $384^3$ resolution whereas the blue dots have $512^3$ resolution, included so as to demonstrate the asymptotic behavior noticed in the cubic system in \cite{Gerhardinger:2022bcw}.  \label{ChangingM}}
 \end{figure*}

While the period-averaged power contained in each multipole highlights the differences between the quartic and cubic dominated systems, we can also study the effects of varying $M$ for a fixed quartic strength, $\xi$. 
Figure \ref{ChangingM} shows how the final quadrupole power depends on the $M \bar{r}$ parameter. 
At $M\bar{r} = 0$, the only source for $\pi$ is $H_{\mu\nu}^2$, but $H_{\mu\nu}$ itself is unsourced. Therefore, any oscillations in $H_{\mu\nu}$ will decay away and, in the long time limit, $\pi$ will approximately satisfy the Klein-Gordon equation.
For large $M$, the power converges to a value that is insensitive to increasing $M$ further, indicating that the UV physics is decoupled from the IR regime.
Comparing Figure \ref{ChangingM} to the analogous one for the cubic only system in \cite{Gerhardinger:2022bcw}, we see that the final quadrupole power depends on the $M\bar{r}$ parameter in a similar way, and we note that the same asymptotic behavior is exhibited with higher resolution runs as was noticed in the previous paper.  
In other words, in the $M\bar{r} >7$ regime, the simulations approximate the $M\bar{r} \rightarrow \infty$, decoupled limit. \\

\begin{figure*}[tbp]
\centering
\includegraphics[width=0.5\textwidth]{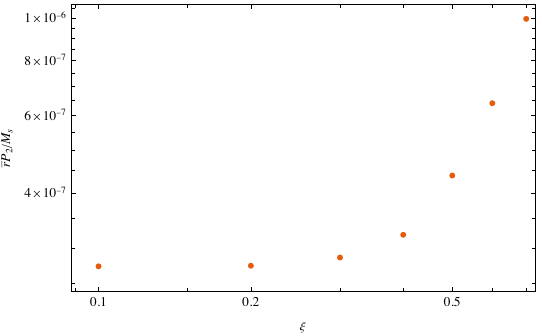}
\caption{Power contained in the quadrupole moment for the fiducial system as a function of the quartic nonlinear coupling strength $\xi$, with fixed $M$ parameter $M \bar{r} = 6$. We note the last point $\xi = 0.7$ reaches convergence although crashes soon after.   \label{ChangingXi}}
 \end{figure*}

We also investigated the $M-\xi$ phase space by examining the effect of increasing the $\xi$ parameter for a single $M$ value on the period averaged quadrupole power. 
Figure \ref{ChangingXi} portrays the final quadrupole power for a fixed mass $M$ value. 
Here, increasing the $\xi$ parameter from 0.0 to 0.7 encodes the progression from the cubic system to non-zero quartic contribution regime, with the quartic term dominating for $\xi \geq 0.6$. This highlights how the power is increasing as we turn on the quartic interaction. \\
 
In the Full Auxiliary Field Method, issues with the boundary conditions meant that we were only able to simulate the system up to a certain $M\bar{r}$ parameter, for a low value of $\xi$, the same limiting value that we found in our previous work with the cubic system \cite{Gerhardinger:2022bcw}. 
Instead of $M\bar{r} = 10$, the cutoff for the cubic Galileon, here the cut off is closer to $M\bar{r} \approx 7$.  
As the quartic term contributes to the equation of motion more and more (i.e. as $\xi$ increases), the cutoff on $M \bar{r}$ decreases, because the cubic contribution at the edge of the box also grows. 
For higher mass runs, the fiducial model in the cubic system could stably evolve for several orbits of the source, but eventually power seeped into the higher frequency modes and caused the simulation to crash.
At the boundary, we calculate derivatives of the auxiliary fields while assuming that the constraints are satisfied -- a good approximation given that the constraints are satisfied exactly and we are far enough away from the source that the $\pi$ field is Klein-Gordon. 
Similarly to the cubic case, it seems that these assumptions are violated for $M \bar{r} \geq 5$, because the cubic terms are non-zero at the edge of the box.  
We have tested the idea that this is a result of the boundary conditions and not the model itself by increasing the spatial resolution of our marginal case $ M \bar{r} = 7$ to effectively move the boundary further from the source, and found that our simulations evolved longer.
In principle, one should be able to probe beyond the range presented here by increasing spatial resolution and the number of points used in the finite derivative equations, although this comes at the cost of computational resources and time (we estimate that on the 20 core machine with 512 GB RAM memory capability, these runs would take at least 6 months to reach stable evolution).

\section{Discussion}

Galileon theories are an interesting class of effective field theories which incorporate the Vainshtein screening mechanism. This mechanism is realized through nonlinear higher derivative interactions which become significant in the region of a massive source. These theories are understood as low energy effective field theories, and are typically not regarded as fundamental beyond some cutoff energy scale. At a practical level, the existence of nonlinear derivative interactions means the classical equations of motion are not well-posed and this renders the numerical evolution strongly sensitive to unstable UV modes. This pathology is of no concern in $S$-matrix calculations where it is well understood how to compute scattering amplitudes in the low energy theory. However it makes extracting physical predictions from Galileon theories difficult. If we want to study and analyze these EFTs numerically we need a system that regulates high momentum modes without disrupting the low energy limit behavior.
In our previous work \cite{Gerhardinger:2022bcw}, we established two different approaches to deal with this. One was to utilize a low-pass filter which automatically switches off any pathological UV behavior. The success of this approach is based on the independence of the long wavelength physics to the precise filtering scale. The second approach was to replace the original Galileon system with a UV completion at the level of the equations of motion, which does not itself have the same high energy problems, but reproduces the same low energy physics. In \cite{Gerhardinger:2022bcw} we showed that both approaches can be successfully used to reproduce the long wavelength behavior expected by analytic analysis. \\

In the present work we extended the UV completion approach to the quartic Galileon system, a system which has proven intransigent to analytic approximations. The low-pass filter method will be considered elsewhere \cite{Giblin:2024}, which is less computationally expensive. Related work \cite{Babichev:2017lrx} and \cite{Bezares:2021yek,Lara:2021piy} use UV completions based on other screening mechanisms considered \cite{Tolley:2009fg,Elder:2014fea,Solomon:2020viz}.  
In this work, we have introduced auxiliary fields that describe new degrees of freedom and obey damped wave equations that regulate the UV behavior. 
These fields trade the nonlinear interaction terms with algebraic functions, thereby making the entire system formally well-posed.
We simulated a binary orbiting system and showed that simulations using this technique are stable. 
Using the same initial data and sources as our previous paper, we demonstrated that the quartic Galileon simulations replicates the previously verified cubic system.
Moreover, we have been able to probe beyond the cubic regime into quartic domination, and in doing so, uncovered the power radiated by each multipole in a quartic domination regime.\\

We encountered the same technical issue in the boundary conditions of massive degrees of freedom which prevented us from simulating large scales of $M$ as we described in \cite{Gerhardinger:2022bcw}.
While we balanced the limiting factors prohibiting further investigations into larger scales (specifically the need for spatial resolution of the source with violating the assumption of negligible contributions to the equation of motion at the boundary), this is a known issue in the treatment of massive outgoing waves (see for instance \cite{Honda:2000gv}).
Hence we expect that further treatment of this numerical issue will allow us to probe into those larger regimes. \\

We have taken a technique developed and verified in \cite{Gerhardinger:2022bcw} and utilized it for a more generic, previously not-able-to-be simulated effective field theory.  
Our hope is that this demonstrates the utility of this technique for numerical work investigating EFTs in general, for instance other extensions to General Relativity and perhaps even beyond cosmology.


\begin{acknowledgments}
The work of MT is supported in part by US Department of Energy (HEP) Award DE-SC0013528. J.T.G.\ is supported in part by the National Science Foundation, PHY-2309919.  The work of AJT is supported by STFC Consolidated Grant ST/T000791/1 and ST/X000575/1. This material is based upon work of MG supported by the U.S. Department of Energy, Office of Science, Office of Advanced Scientific Computing Research, Department of Energy Computational Science Graduate Fellowship under Award Number DE-SC0023112. We acknowledge the National Science Foundation, Kenyon College and the Kenyon College Department of Physics for providing the hardware used to carry out these simulations.\\

Disclaimer:``This report was prepared as an account of work sponsored by an agency of the United States Government. Neither the United States Government nor any agency thereof, nor any of their employees, makes any warranty, express or implied, or assumes any legal liability or responsibility for the accuracy, completeness, or usefulness of any information, apparatus, product, or process disclosed, or represents that its use would not infringe privately owned rights. Reference herein to any specific commercial product, process, or service by trade name, trademark, manufacturer, or otherwise does not necessarily constitute or imply its endorsement, recommendation, or favoring by the United States Government or any agency thereof. The views and opinions of authors expressed herein do not necessarily state or reflect those of the United States Government or any agency thereof."

\end{acknowledgments}

\bibliographystyle{apsrev4-1} 
\bibliography{references.bib}

\end{document}